\documentclass[%
 reprint,
 amsmath,amssymb,
 aps,
]{revtex4-1}

\usepackage{graphicx}
\usepackage{dcolumn}
\usepackage{bm}

\begin{document}

\preprint{APS/123-QED}

\title{Vortex stream generation and enhanced propagation in a polariton superfluid}

\author{Giovanni Lerario}%
 \email{giovanni.lerario@lkb.upmc.fr}
\author{Anne Ma\^{i}tre}
\author{Rajiv Boddeda}

\author{Quentin Glorieux}%
\author{Elisabeth Giacobino}%
\author{Simon Pigeon}%
\author{Alberto Bramati}%
 \email{alberto.bramati@lkb.upmc.fr}
\affiliation{%
 Laboratoire Kastler Brossel, Sorbonne Universit\'{e}, CNRS, ENS-Universit\'{e} PSL, Coll\`{e}ge de France, Paris 75005, France
}%

\date{\today}

\begin{abstract}
In this work, we implement a new experimental configuration which exploits the specific properties of the optical bistability exhibited by the polariton system and we demonstrate the generation of a superfluid turbulent flow in the wake of a potential barrier. The propagation and direction of the turbulent flow are sustained by a support beam on distances an order of magnitude longer than previously reported. This novel technique is a powerful tool for the controlled generation and propagation of quantum turbulences and paves the way to the study of the hydrodynamic of quantum turbulence in driven-dissipative 2D polariton systems.
\end{abstract}

\maketitle

\section{Introduction}

Exciton-polaritons are bosonic quasi-particles arising from the strong coupling between excitons and photons. They have hybrid properties carried by their bare components, and exhibit highly non-linear behavior due to their excitonic content \cite{Vladimirova2010}. Since the first demonstration of Bose Einstein condensation in exciton-polariton microcavities \cite{Kasprzak2006b}, these systems became an attractive platform to study the hydrodynamics of quantum fluids.  Thanks to the ability to prepare a polariton superfluid steady flow, the hydrodynamic nucleation of quantum turbulences was reported \cite{Amo2009, Roumpos2011, Nardin2011, Amo2011}.

In the last decade, many efforts have been made to control and manipulate the generation and propagation of vortices in polariton systems \cite{Sanvitto2011, Dall2014, Boulier2016a, Hivet2014}. The vortices in a superfluid are collective excitations formed by a low density core with quantized circulation around it, which makes them resilient to spontaneous decay (i.e. they are topologically protected). Since they are the fundamental block of the  turbulence in superfluids, the study of their dynamics raises a great interest.

Moreover, optical vortices are extremely promising for their application in quantum optical information science. In optics, modes carrying orbital angular momentum are used to encode and transport information \cite{Willner2015}. Quantum vortices in planar microcavities are ideal candidates for binary encoding (thanks to their robustness to the perturbations and their clockwise/counter-clockwise phase winding) towards on-chip technology. Furthermore, orbital angular momentum entanglement is at the forefront of quantum information science \cite{Mair2001, Hiesmayr2016, Fickler2016, Nagali2009, Giovannini2011, Marrucci2011}.

Different experimental configurations have been used to generate polariton vortices. In early experiments, quantized vortices have been observed in a condensate pumped with a continuous-wave (cw) non-resonant excitation \cite{Lagoudakis2008}, spontaneously emerging from the overlap between the condensate and a disordered potential landscape.  Quantized vortices pairs have been also obtained using pulsed resonant pumping \cite{Nardin2011, Lerario2017a}. In this case, because of the dissipative nature of exciton-polaritons, the vortices lifetime is limited by the polariton lifetime, thus limiting the study of their dynamics. With a cw resonant pumping, in order to ensure the absence of phase fixing and to allow the appearance of topological excitations, many experiments were performed in the free propagation regime, outside of the excitation area by the laser. However, also in this configuration, the polariton density and the propagation distance of the quantum fluid are limited by the polariton lifetime  \cite{Amo2011}.
Thereafter, cw resonant and non-resonant pumping have been used to generate stationary vortices with in-plane engineering of the potential landscapes or even in Optical Parametric Oscillation configuration \cite{Sanvitto2011, Dall2014, Boulier2016a, Hivet2014}. However, all these cw configurations inhibit vortices free propagation–the vortices core is spatially pinned by the excitation intensity shape–and impose strong constraints on their free interaction.

A quasi-resonantly driven semiconductor microcavity  exhibits a bistable behavior when the pump laser is slightly blue detuned with respect to the polariton resonance \cite{Baas2004}. The optical bistability, theoretically illustrated in figure \ref{fig:fig1}a, has been used in polariton systems for switching on and off the polariton population by triggering it with a weak control beam, allowing the achievement of polariton switches and logic gates \cite{DeGiorgi2012, Ballarini2013}. Moreover, the upper branch of the bistability loop, where the polariton density and the nonlinear interactions are very strong, is closely linked to striking phenomena such as superfluidity and supersonic Cherenkov propagation \cite{Amo2009}.

In this work, we consider the bistable hysteresis loop, highlighted in blue in Fig \ref{fig:fig1}a, comprised between the lower and upper turning points of the bistability curve, where the system can be either in a linear or nonlinear regime. Remarkably, as shown in a recent theoretical proposal \cite{Pigeon2017}, in this region the phase of the fluid is not fixed by the resonant driving field but is instead free to evolve and at the same time, the polariton density can be high enough to observe spontaneous collective superfluid behaviour mediated by the nonlinearity of the system.

We demonstrate for the first time that in the bistable regime, vortex-antivortex pairs can be generated and controlled while their propagation is strongly enhanced far beyond the polariton free propagation length. To experimentally implement this configuration, we make use of a set-up with two beams: a support beam on a large area with an intensity inside the bistability cycle and a localized seed beam with an intensity above the bistability regime that creates a superfluid

flow and triggers the support into the upper branch of the bistability cycle (see Fig.\ref{fig:fig1}a and \ref{fig:fig1}b). The support beam  then sustains the superfluid propagation along the microcavity plane (see figure \ref{fig:fig1}b). Placing an obstacle in the supported region, we show that topological excitations can be hydrodynamically generated in the wake of the obstacle and propagate as long as the polariton superfluid remains in the bistable regime \cite{Pigeon2017}.

Furthermore, we show that the vortex propagation direction can be controlled by changing the in-plane wavevector of the support beam. Finally, modifying the support intensity, we can also control the density of vortices in the turbulent stream. The propagation distance appears to be only limited by the experimental need of a large homogeneous support beam, offering in principle limitless propagation of a quantum turbulent flow.

\section{Results and Discussion}

The polariton microcavity under investigation is made of three $In_{0.04}Ga_{0.96}$ As/GaAs quantum wells embedded between two GaAs/AlAs-based DBRs, 21 pairs for the front DBR and 24 pairs for the back one. The measurements are performed at cryogenic temperature and in transmission configuration.

\begin{figure}[t]
\includegraphics[width=0.49\textwidth]{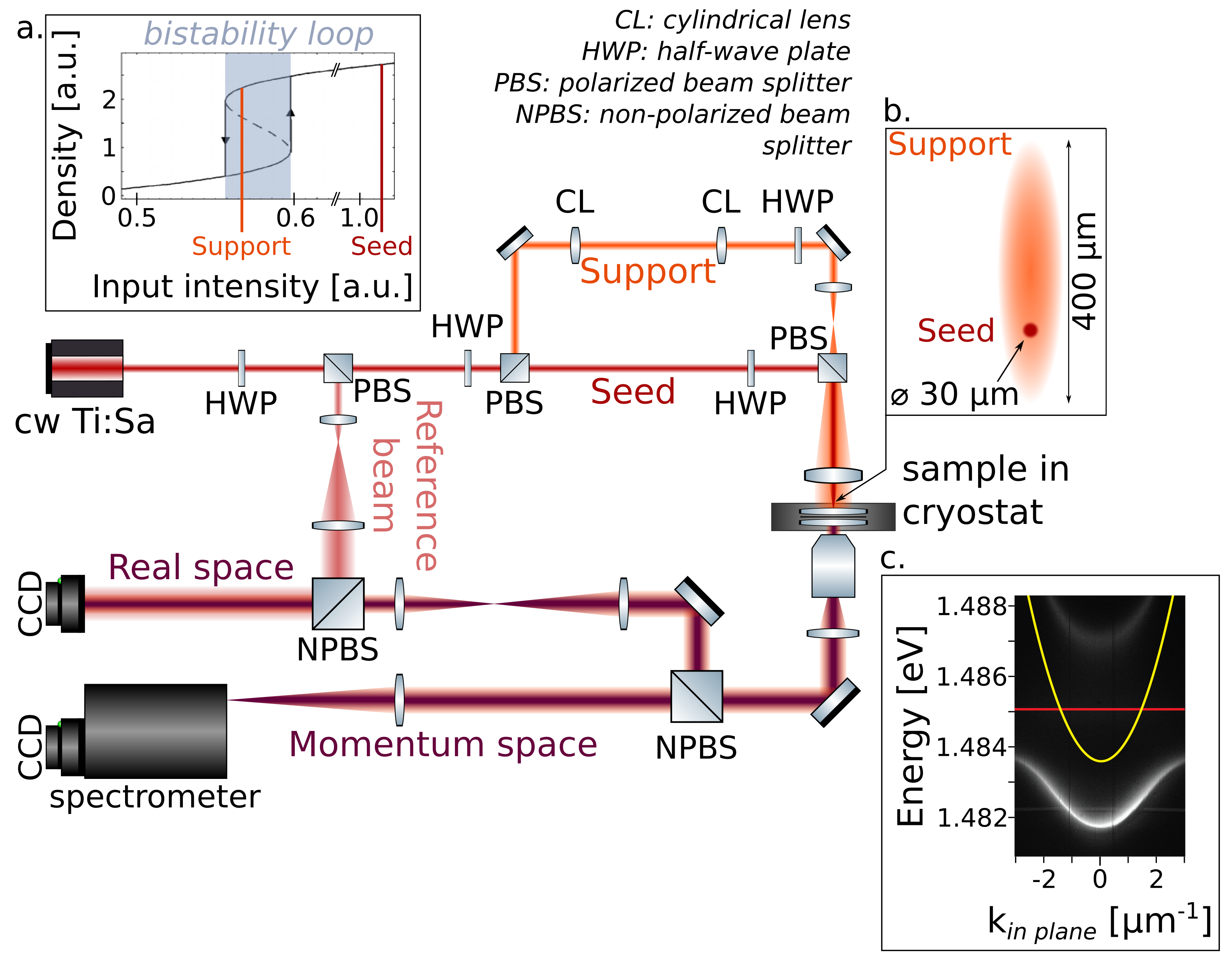}
\caption{\label{fig:fig1} Bistability, setup and polariton dispersion.
Inset (a) illustrates the positions of the support and seed beams with respect to the theoretical bistability curve. The support intensity insures the fluid to stay inside the bistability loop, while the seed beam locks it on the upper branch. The highlighted region corresponds to the bistability loop, where the phase constraint imposed by the driving field is released.\\
 Sketch of the optical setup. The excitation line consists of a support beam path (collimated on the sample surface) and a seed beam path (focused on the sample surface). The two beams are superposed at the sample surface. Using two cylindrical lenses in the excitation path, the support beam has an elliptic shape. Inset (b) shows the configuration of the seed and support on the sample surface. The detection line consists of two different paths projecting the real space and momentum space maps on two CCD cameras (see main text for further details). c) Polariton dispersion under non-resonant excitation (2.54 eV CW laser). The strong coupling between the exciton transition (red line) and the photon mode (yellow line) gives rise to the upper and lower polariton branches.
}
\end{figure}

The optical setup is sketched in Figure \ref{fig:fig1}. A cw single mode Titane-Sapphire laser at 831nm pumps the polaritons in the quasi-resonant excitation configuration. A first half wave plate (HWP) and polarizing beam splitter (PBS) extract a reference beam for the homodyne interferometric detection. The main beam is split in two different channels using a second HPW and  PBS, one channel for the support beam and the other one for the seed beam. The seed beam is focused on the sample surface in an area with FWHM = 30 $ \mu$m and intensity $\rho_{p} $= 10.6 W/mm$^{2}$. Making use of two cylindrical lenses (CL) along the support excitation line, the support beam intensity has an elliptical shape and it is collimated (100x400 $\mu $m FWHM) at the sample surface in order to cover a large area of the sample. The support input intensity is 5.8 W/mm$^{2}$. The two beams merge in the same excitation channel through a PBS, generating a seed spot focused near the edge of the large area illuminated by the support beam (Figure \ref{fig:fig1}b).

\begin{figure*}[t]
\includegraphics[width=0.99\textwidth]{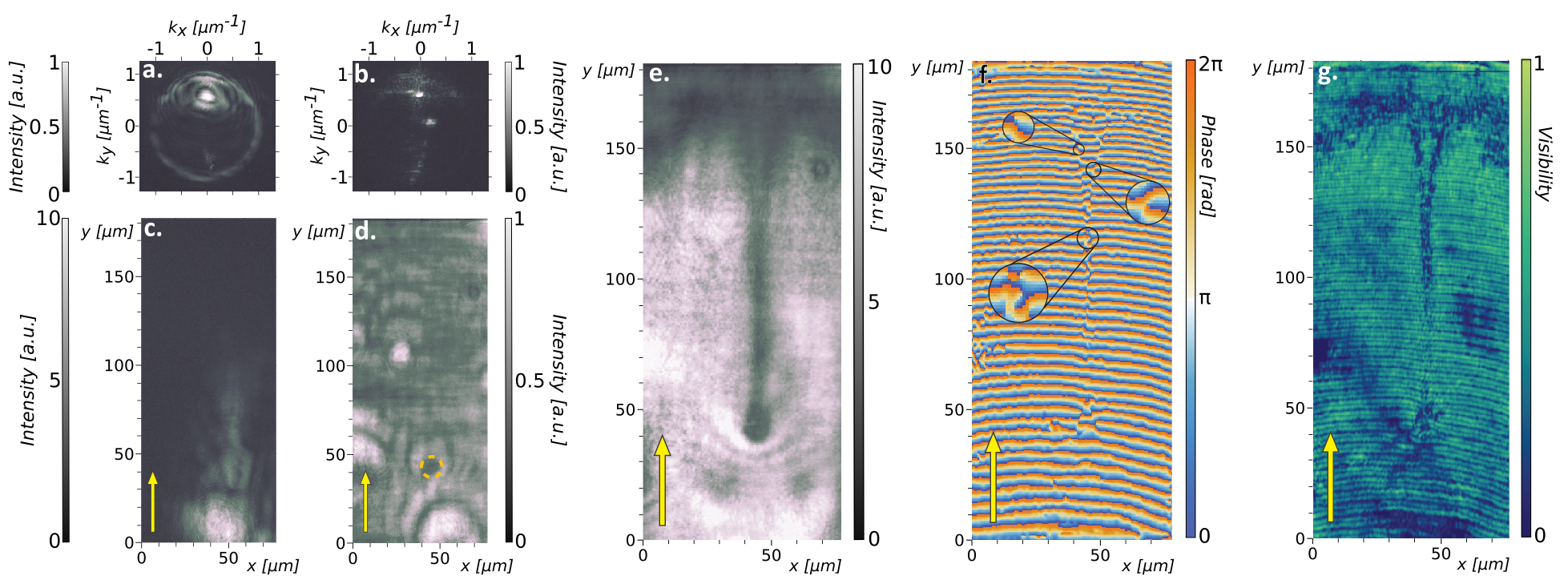}
\caption{\label{fig:fig2} Vortex generation and propagation. 
Momentum maps of seed-only (a) and support-only (b). The excitation detuning is +0.26 meV for both the beams. c) and d) panels show the space intensity distribution of the seed-only and support-only, respectively. The arrows indicate the flow direction. The position of the obstacle in (d) is shown by the yellow dashed circle. Note that for the support-only (d) the intensity scale is divided by a factor of ten. Intensity (e), phase (f) and visibility (g) maps when the support and the seed simultaneously excite the sample. Thanks to the presence of the support beam, the polariton population is drastically increased in a wide area. The vortices flowing in the wake of the defect generate a long dark shadow in the fluid propagation direction. The time integration over a continuous flow of vortices induces a blurring of the phase (lower zoom in panel f) along the vortex stream and a consequent decrease of the visibility. Fork-shaped patterns (central zoom in panel f) demonstrate the formation of localized vortices and the upper zoom in panel f indicates the formation of unstable grey solitons at the border of the locked area (see main text for further details). Panels e,f,g are plotted keeping fixed the real space image orientation.
}
\end{figure*}

We can control the polariton group velocity, or flow velocity, by varying the angle of incidence $\theta $ of the pump with respect to the device plane ($ k = k_{0} \sin(\theta) $ , with $k_{0} $ the wavevector of the pumping laser). The detection channel is divided into two different paths by a beam splitter. The real space image of the sample surface is detected with a CCD camera in one arm, giving the map of the the polariton fluid density, while a CCD camera placed in the second detection arm images the momentum space map. By using a spectrometer before the CCD camera in this second arm, the full dispersion curves can be measured by pumping the system with a non-resonant excitation (2.54 eV cw laser). Fig \ref{fig:fig1}c shows the dispersion curves corresponding to the upper and lower polaritons arising from the strong coupling between the exciton level (red line) and the planar microcavity photon mode (yellow line). The cavity has a negative exciton-photon detuning (-1.4 meV), the Rabi splitting is 5.2 meV and the polariton lifetime is 14 ps. The experiments are operated close to resonance with the lower polariton mode.

Figures \ref{fig:fig2}a and \ref{fig:fig2}b show the momentum space maps in the quasi-resonant excitation configuration for seed-only and support-only, respectively. Both excitation beams have the same wavevector ($|k_{y}|= 0.6$ $\mu m^{-1} $ and $|k_{x}|= 0$ $\mu m^{-1}$). The flow speed depends on the polariton dispersion according to the formula $ v_{flow}  = \dfrac{1}{\hbar} \dfrac{\partial E}{\partial k} $. In our case, the flow speed is 0.9 $\mu $m/ps with the flow direction along the support beam major axis.

The excitation energy is close to resonance with the lower polariton branch at low excitation power. The scattering ring in Figure 2a ($|k|= 1$ $\mu m^{-1}$ radius) highlights the detuning of the excitation with respect to the lower polariton resonance, which is of 0.26 meV at the excitation in-plane wavevector. Unsing a small the excitation spot in the momentum space and increasing the excitation power, the blueshift of the polariton dispersion induced by to the polariton-polariton repulsive interactions allows for the abrupt increase of the polariton density when the system locks to the excitation laser energy. With a blue detuned driving field, one can then observe a bistability cycle \cite{Baas2004,Pigeon2017} as for any Kerr medium in a microresonator.

In the case of support-only (Fig \ref{fig:fig2}d), the system lies in the lower part of the bistability loop (note that in Fig \ref{fig:fig2}d the support-only intensity map, the scale is divided by a factor of ten with respect to the other intensity maps) and few photons enter into the microcavity. The polariton flow propagation in the disordered random potential of the microcavity leads to interference effects offering bright and dark spots. One dark spot, indicated by the orange dashed circle in Fig \ref{fig:fig2}d, is noticeable by its size, $ \approx$ 8 $\mu$m  diameter, and corresponds to a high potential barrier, which will create vortices downstream.

When the seed beam is superposed with the support beam (Figure \ref{fig:fig2}e), the local increase of the polariton-polariton interactions blue shifts the mode resonance, allowing the jump of the system to the upper part of the bistability loop, together with  a drastic increase of the polariton density. The high polariton density region extends over a much wider area than where the seed beam is located, and it is only limited by the support size. This is possible because the intensity of the  support beam is carefully adjusted to stay within the hysteresis loop of the bistability curve.

The sound velocity depends on the polariton density, 
$ c_{sound}  = \sqrt{ \dfrac{\hbar g |\psi|^{2} }{m_{pol}} } $
, with $g$ the polariton-polariton interaction constant, $m_{pol} $ the polariton mass and $\psi $ the polariton wave-function. $c_{sound} $ is 0.78 $\mu$m/ps in the area where the support beams locks to the upper part of the bistability cycle. The flow velocity (0.9 $\mu $m/ps, as mentioned above) is then larger than the speed of sound and the propagation is supersonic, allowing for the generation of Cherenkov wavefronts upstream the obstacle and for the propagation of perturbations downstream. Here, a long dark shadow appears in the wake of the defect along the support propagation direction. The presence of a potential barrier on the polariton flow leads to the generation of vortex-antivortex pairs. The time integration over a flowing stream of vortex pairs continuously generated at the defect position leads to a shadow in the wake of the defect, as predicted in Ref. \cite{Pigeon2017} and reported here. Because of the time integration, the intensity of the shadow in the wake of the defect depends on the vortex generation rate: higher is the generation rate, darker is the shadow. According to  the theoretical predictions, the vortex generation rate is maximal close to the lower limit of the bistability loop, which allows us to maximize the contrast and to clearly observe the turbulence stream. The vortices propagate for more than 120 $\mu$m, an order of magnitude longer than the distance supported by the polariton lifetime, that is 12 $\mu$m. This propagation distance is only limited by the support field profile which needs to be as flat as possible.

\begin{figure}[t]
\includegraphics[width=0.49\textwidth]{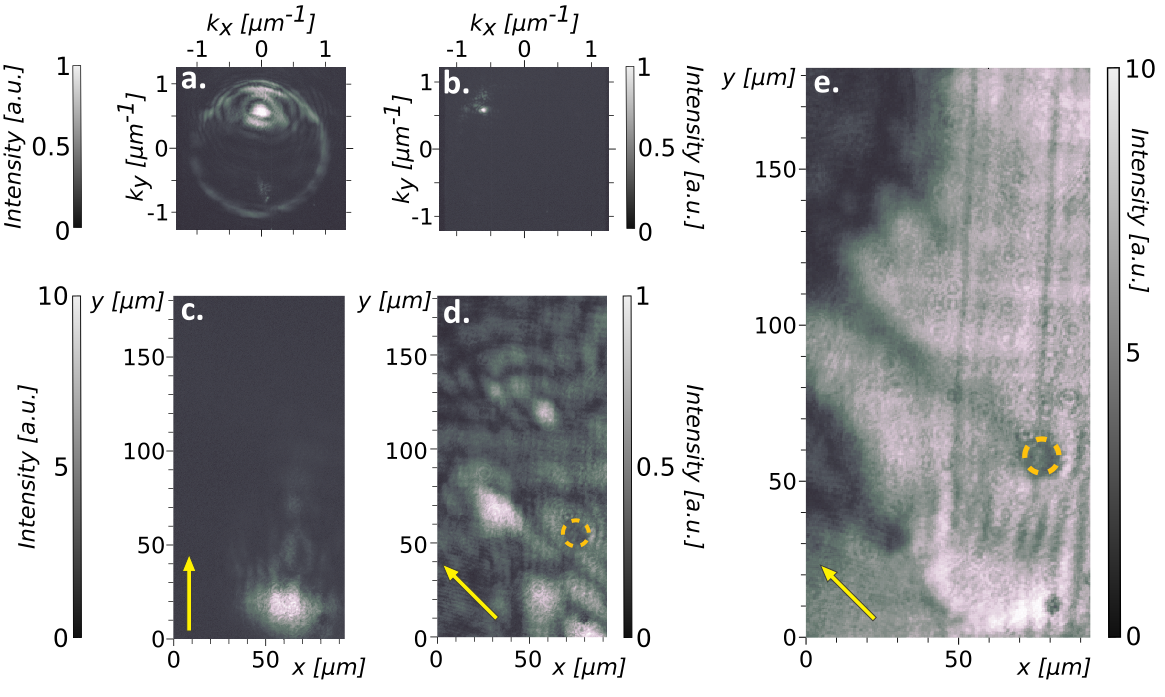}
\caption{\label{fig:fig3} Control of the vortex propagation direction. 
Momentum maps of seed-only (a) and support-only (b). The excitation detuning is +0.26 meV for both the beams, but they are at an angle of about 40$^{\circ} $ with respect to each other. The real space images of the seed-only (c) and support-only (d) intensity show the two different propagation directions. The arrows indicate the flow direction. Note that the support-only intensity scale is divided by a factor of ten. Panel (e) shows the intensity map when the support and seed simultaneously excite the sample. The vortex stream flows in the direction fixed by the support beam and propagates for 90 $\mu$m. 
}
\end{figure}

To further justify that the observed shadow indeed results from the time averaging of a quantum turbulent flow, the phase pattern of the fluid is analyzed. A homodyne interferometer is used to probe the phase pattern associated to Fig \ref{fig:fig2}e. A portion of the support beam is extracted along the excitation line and used as a reference beam with a homogeneous phase, making it interfere on the CCD camera with the signal coming from the microcavity. The time integration over a continuous flow of the vortex stream results in the blurring of the phase in the wake of the defect. Indeed, the lower inset of the interferogram (Fig \ref{fig:fig2}f) presents such a blurred phase at a position coinciding with the shadow stream. To enlighten this phenomenon we report in Figure \ref{fig:fig2}g the corresponding visibility which decreases along the turbulent stream line. Fork-shaped patterns associated to localized vortices along the flow are also visible in the central zoom in Fig \ref{fig:fig2}f. At the border of the locked area the polariton density is decreasing because of the Gaussian shape of the support beam. Therefore  in this region the vortex core radius increases due to the increase of the healing length. This effect induces the merging of the ejected vortices and the formation of unstable grey solitons (upper zoom in Fig \ref{fig:fig2}f), which repel each other and generate the Y-shape observed in the experiment. This Y shape has also been observed for solitons in previous experiments \cite{Amo2011}.

\begin{figure}[b]
\includegraphics[width=0.49\textwidth]{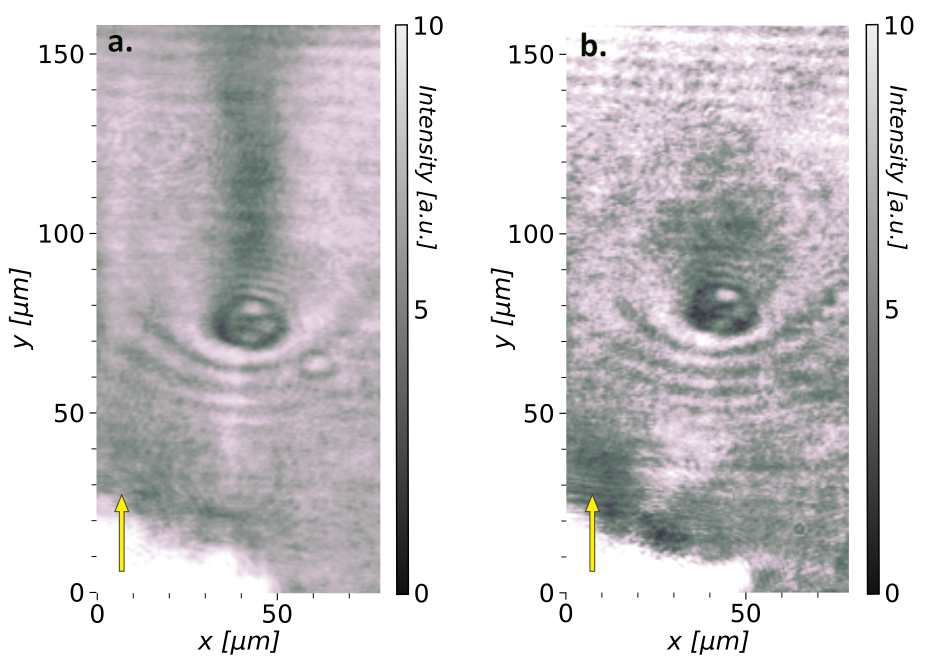}
\caption{\label{fig:fig4} Vortices suppression. 
a) Vortex stream generation and propagation in the wake of the defect for 5.8 W/mm$^{2}$ support power density. b) Increasing the support intensity up to 6 W/mm$^{2}$, the phase of the support is strongly imprinted on the fluid and, consequently, the vortex generation is inhibited.
}
\end{figure}

Moreover, the seed-support configuration allows to control the vortex propagation direction via the support wavevector orientation. Figure \ref{fig:fig3}e shows the stream of vortices flowing with a non-zero component in $k_{x}$. Here, the seed momentum is $|k_{y}|= 0.6$ $\mu m^{-1} $ (and $|k_{x}|= 0$ $\mu m^{-1}$, Figure \ref{fig:fig3}a),  while the support momentum orientation, with the same module, is displaced at about 40$^{\circ}$ with respect to the seed one (Figure \ref{fig:fig3}b). Panels \ref{fig:fig3}c and \ref{fig:fig3}d show the seed-only and support-only space maps, respectively. As evident from Figure \ref{fig:fig2} and \ref{fig:fig3}, the vortices stream flows with a propagation direction defined by the support beam direction.

As specified above, the intensity of the shadow depends on the vortex generation rate in time-integrated measurements. By increasing the support density above the bistable regime the vortex generation is strongly inhibited  because the fluid phase is fixed to the support beam phase. Experimentally, this translates into the suppression of the shadow in the wake of the defect when increasing the support power as predicted in Ref. \cite{Pigeon2017}. In Figure \ref{fig:fig4}a we can observe the vortex stream propagating in the wake of the defect; here, the experimental configuration and parameters are identical to the ones previously described. When the excitation density is increased to 6 W/mm$^{2}$ (Figure \ref{fig:fig4}b), which is just above the limit of the bistable regime, the vortex stream is completely suppressed by the phase constraint imposed by the support beam, with almost no change in the fluid properties (speed and density).

\section{Conclusions}

In conclusion, we demonstrated the generation of a vortex stream flowing over more than 120 $\mu$m in the plane of a polariton microcavity, implementing an original method proposed by Pigeon et al. \cite{Pigeon2017}. In this approach we used the bistability to release the phase fixing of the polariton fluid to the resonant support beam, while maintaining the particle density sufficiently high to observe topological excitations. Vortices can then be sustained without being limited by the polariton lifetime and can flow over macroscopic distances. Additionally, we demonstrated that the vortex stream direction can be controlled by the orientation of the support beam wavevector and the vortex density can be tuned varying the support intensity within the bistability cycle. This is the first implementation of an experimental configuration that allows for the controlled propagation and interaction of polariton vortices, paving the way for the on-chip macroscopic propagation and efficient manipulation of topological excitations.

\section{Acknowledgments}

This work has received funding from the French ANR grants (C-FLigHT 138678 and QFL) and from the European Union Horizon 2020 research and innovation programme under grant agreement No 820392 (PhoQuS). QG and AB thank the Institut Universitaire de France (IUF) for support.

\nocite{*}

\bibliography{LKB-bibs-bib_VortexStream_Gianni}

\end{document}